\documentclass[aps,onecolumn,longbibliography]{revtex4-2}
\usepackage[dvips]{graphics,graphicx}
\usepackage{amsmath}
\usepackage[usenames, dvipsnames]{color}
\usepackage{graphicx}
\usepackage{graphics}
\usepackage[outdir=./]{epstopdf}

\begin{document}
\title{Quantum manifestation of the classical bifurcation in the driven dissipative Bose-Hubbard dimer}
\author{ P. S. Muraev$^{1,2}$, D. N. Maksimov$^{1,3}$, and A. R. Kolovsky$^{1,2}$}
\affiliation{$^1$Kirensky Institute of Physics, Federal Research Center KSC SB RAS, 660036, Krasnoyarsk, Russia}
\affiliation{$^2$School of Engineering Physics and Radio Electronics, Siberian Federal University, 660041, Krasnoyarsk, Russia}
\affiliation{$^3$IRC SQC, Siberian Federal University, 660041, Krasnoyarsk, Russia}
\date{\today}
\begin{abstract}
We analyze the classical and quantum dynamics of the driven dissipative Bose-Hubbard dimer. Under variation of the driving frequency, the classical system is shown to exhibit a bifurcation to the limit cycle, where its steady-state solution corresponds to periodic oscillation with the frequency unrelated to the driving frequency. This bifurcation is shown to lead to a peculiarity in the \emph{stationary} single-particle density matrix of the quantum system. The case of the Bose-Hubbard trimer, where the discussed limit cycle bifurcates into a chaotic attractor, is briefly discussed.
\end{abstract}
\maketitle

\section{Introduction}

\begin{figure}
\includegraphics[width=10.5 cm]{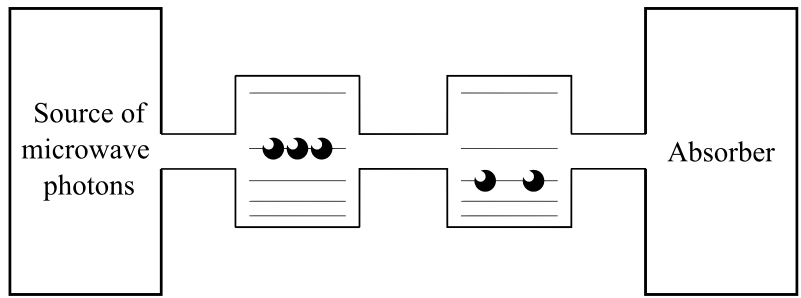}
\caption{Pictorial presentation of the considered model. The figure shows two coupled nonlinear micro-cavities with three photons in the first cavity and two photons in the second cavity. It is assumed that the microwave photons are injected into the first cavity by using a generator of the microwave field, and the transmitted photons are absorbed with some probability by a measurement device.\label{Fig0}}
\end{figure}

In the present work, we analyze the dynamics of the two-site driven dissipative Bose-Hubbard (BH) model.  Similar to the conservative two-site BH model, which provides a model for the Josephson oscillations and the phenomenon of self-trapping \cite{Smerzi_1997, Albiez_2005, Levy_2007, Abbarchi_2013}, the driven dissipative BH systems model a number of phenomena in open quantum systems. For example, the one-site system, which is nothing else than the driven dissipative nonlinear oscillators, is the paradigm system for quantum bistability (see Ref.\cite{kolovsky2020bistability} and references therein). Extending the system to two sites enriches its dynamics and drastically complicates the classical bifurcation diagram \cite{Giraldo_2020, Giraldo_2022}, which poses the problem of a quantum signature of these bifurcations \cite{Lled__2019, Lled__2020}. Finally, the classical three-site BH system can show the chaotic attractor that brings us to the problem of the dissipative Quantum Chaos \cite{braun2001dissipative, S__2020}. We mention that nowadays the few-site open BH model can be and has been realized experimentally by using different physical platforms, among which the most successful are exciton-polariton semi-conductor systems \cite{Lagoudakis_2010, Abbarchi_2013, Rodriguez_2016} and super-conducting circuits \cite{Eichler_2014, Raftery_2014, Fedorov_2021}. In what follows, we theoretically analyze the two- and three-site driven dissipative BH model by keeping in mind laboratory experiments with the chain of transmons, which are micro-cavities coupled to Josephson's junctions. The presence of Josephson's junction introduces an effective inter-particle interaction for photons in the cavity and, thus, each transmon can be viewed as a quantum nonlinear oscillator, see Fig.\ref{Fig0}. We mention that in the present work we do not try to relate the model parameters to the system parameters used in one or the other laboratory experiment. In this sense, the model depicted in Fig.\ref{Fig0} captures only the general scheme of these experiments, where quantum nonlinear oscillators are arranged in the `transmission line’ and one measures the amplitude of the transmitted signal, i.e., the current of the microwave photons. 

\section{Quantum dynamics}
\label{sec2}

We consider two coupled transmons where the first transmon is excited by a microwave generator and the transmitted signal is read from the second transmon. The governing equation for the system density matrix $\widehat{{\cal R}}$ reads

\begin{equation}\label{M_Eq}
\frac{\partial \widehat{{\cal R}}}{\partial t}=-\frac{i}{\hbar}[\widehat{{\cal H}}, \widehat{{\cal R}}] -\frac{\gamma}{2}
\left(\hat{a}_2^{\dagger}\hat{a}_2\widehat{\cal R }-2\hat{a}_2\widehat{\cal R }\hat{a}_2^{\dagger}
+\widehat{\cal R }\hat{a}_2^{\dagger}\hat{a}_2 \right) \;,
\end{equation}
where $\gamma$ is the decay constant proportional to the absorption rate.  Using the rotating wave approximation, the Hamiltonian $\widehat{{\cal H}}$ in Eq.~(\ref{M_Eq}) has the form 

\begin{equation}\label{q_Hamiltonian}
\widehat{{\cal H}}= -\hbar \Delta \sum_{\ell=1}^{2}\hat{n}_{\ell}
-\frac{\hbar J}{2}\left( \hat{a}_{2}^{\dagger}\hat{a}_{1} +{\rm h.c.} \right) 
+\frac{\hbar^2 U}{2}\sum_{\ell=1}^{2}\hat{n}_{\ell} (\hat{n}_{\ell}-1) + \frac{\sqrt{\hbar}\Omega}{2}(\hat{a}_1^\dagger + \hat{a}_1) \;,
\end{equation}
where $\hat{a}^\dagger$ and $\hat{a}$ are the creation and annihilation operators with the commutation relation $[\hat{a},\hat{a}^\dagger]=1$, $\hat{n}_{\ell}$ is the number operator, $\Omega$ is the Rabi frequency, $\Delta$ the detuning defined as the difference between the driving frequency and the cavity eigen-frequency, $J$ the coupling constant, $U$ the microscopic interaction constant, and $\hbar$ is the dimensionless Planck constant which determines how close is the system to its classical counterpart. For quantum systems with the conserved number of particles $N$, one can define the dimensionless  Planck constant as $\hbar=1/N$. In our case, where the number of particles is not conserved, $\hbar$ is just the scaling parameter that leaves invariant the classical dynamics.  We focus on the case $\hbar\ll 1$ where the quantum dynamics shows similarities with the classical dynamics. In the opposite limit $\hbar>1$ the system dynamics is dominated by the multi-photon resonances which have no classical analog.

Our main object of interest is  the single-particle density matrix  (SPDM) $\hat{\rho}$ which is defined as follows,

\begin{equation}
\label{3}
\rho_{\ell,m}(t)={\rm Tr}[\hat{a}^\dagger_\ell \hat{a}_m \widehat{{\cal R}}(t)] .
\end{equation}

\begin{figure}
\includegraphics[width=10.5 cm]{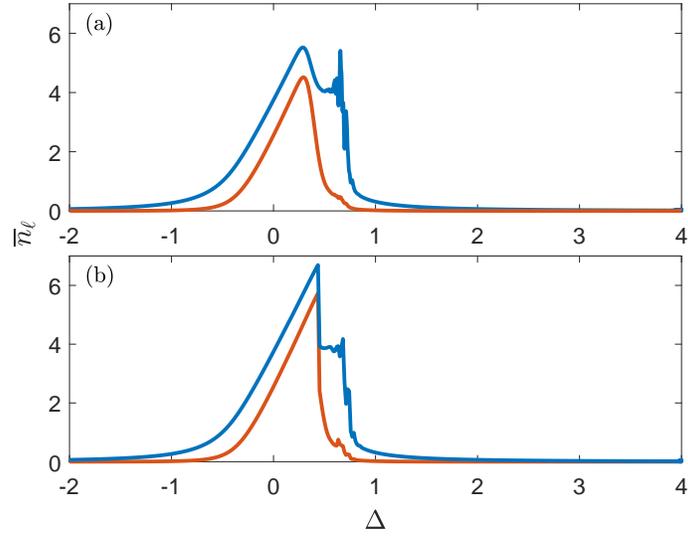}
\caption{The mean number of bosons in the BH dimer as the function of the detuning $\Delta$. The blue and red lines correspond to the first and second sites, respectively. (\textbf{a})  Quasi-adiabatic passage with the sweeping rate $d\Delta/dt=0.0012$. (\textbf{b}) Steady-state solution for different  $\Delta$. The  system parameters are $J = 0.5$, $U = 0.5$, $\Omega = 0.5$, $\hbar = 1\slash 4$, $\gamma = 0.2$.
\label{Fig1}}
\end{figure}

The diagonal elements of the SPDM obviously determine the populations of the sites, while off-diagonal elements determine the current of Bose particles (photons) between the sites. We find the stationary SPDM for different $\Delta$, which will be our control parameter, by using two methods: (i) by evolving the system for fixed $\Delta$ for a long time sufficient to reach the steady-state regime, and (ii) by sweeping $\Delta$ in the {\em negative} direction starting from a large positive $\Delta$. In both cases, the initial condition corresponds to the empty system, i.e., $\hat{\rho}(t=0)=0$. The obtained results are depicted in Fig.~\ref{Fig1}. It is seen in Fig.~\ref{Fig1} that stationary occupations of the chain sites (i.e., the mean number of photons in transmons) show a kind of plateau in the certain interval of $\Delta$.  Notice that this plateau is absent for the one-site BH model. It is argued in the next section that this peculiarity is a signature of the attractor bifurcation which one finds in the classical counterpart of the system (\ref{q_Hamiltonian}).

\section{Classical dynamics}
\label{sec3}

\begin{figure}
\includegraphics[width=10.5 cm]{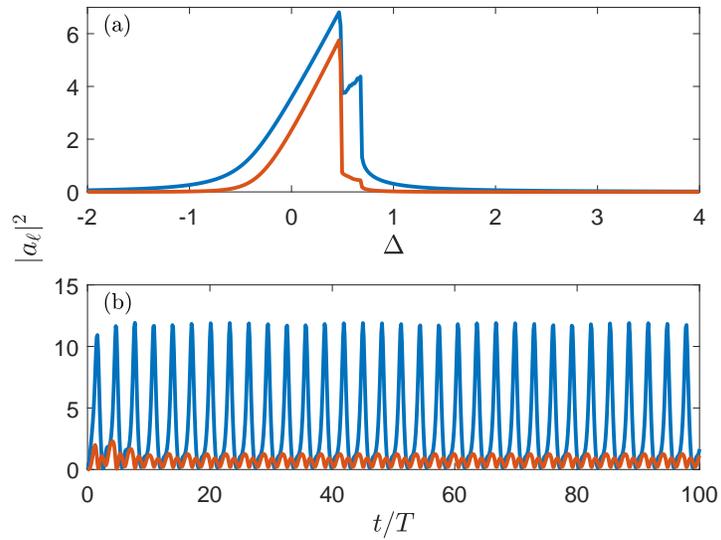}
\caption{(\textbf{a}) The mean number of bosons in the BH dimer averaged over time as a function of detuning $\Delta$.  (\textbf{b}) System dynamics for $\Delta=0.5$. Shown are the squared amplitudes of the local oscillators as the function of time.}
\label{Fig2}
\end{figure}

The classical (mean-field) dynamics of the system is governed by the equations

\begin{align}\label{4}
 & i\dot{a}_1=(-\Delta+U|a_1|^2)a_1-\frac{J}{2}a_2+ \frac{\Omega}{2} \nonumber \\
    & i\dot{a}_2=(-\Delta+U|a_2|^2)a_2-\frac{J}{2}a_1-i\frac{\gamma}{2}a_2
\end{align}
where $a_\ell$ are complex amplitudes of the local oscillators. The numerical simulations were performed by using the fourth-order Runge-Kutta method. 
It is found that for nonzero $\gamma$ the system (\ref{4}) relaxes in course of time to some attractor which determines the system's stationary response to the external driving. In what follows we shall be interested only in attractors whose basin contains the point ${\bf a}=0$. For $\Delta<0.48$ and $\Delta>0.77$ we found  these attractors to be simple focuses (in the rotating frame), where the populations $|a_\ell (t)|^2$ approach their stationary values depicted  in Fig.~\ref{Fig2}(a) by the blue and red solid lines. However, in the interval $0.48 < \Delta < 0.77$ these simple attractors bifurcate into the limit cycle, where the steady state solution of Eq.~(\ref{4}) corresponds to periodic oscillations of the oscillator amplitudes with the frequency $\nu=\nu(\Delta)$ not related to the driving frequency, see Fig.~\ref{Fig2}(b). \footnote{ Bifurcation of a simple attractor into a limit cycle in the driven dissipative two-site BH system was discussed earlier in Ref.\cite{Lled__2019, Lled__2020} where the authors considered a specific model with {\em nonlocal} driving and dissipation. Also, when addressing the quantum system, the authors focussed on the transient dynamics for particular initial conditions but not on the stationary regime.} Besides this characteristic frequency, we also introduce the mean squared amplitudes $\overline{|a_\ell|^2}$ where the bar denotes the time average. We depict these quantities in  Fig.~\ref{Fig2}(a) by using the same line styles.  Comparing now Fig.~\ref{Fig1}(b) and Fig.~\ref{Fig2}(a) we conclude that bifurcation of the classical attractor is well reflected in the quantum dynamics of the BH dimer already for $\hbar=1/4$.

\section{Dissipative BH trimer}
\label{sec4}

\begin{figure}
\includegraphics[width=10.5 cm]{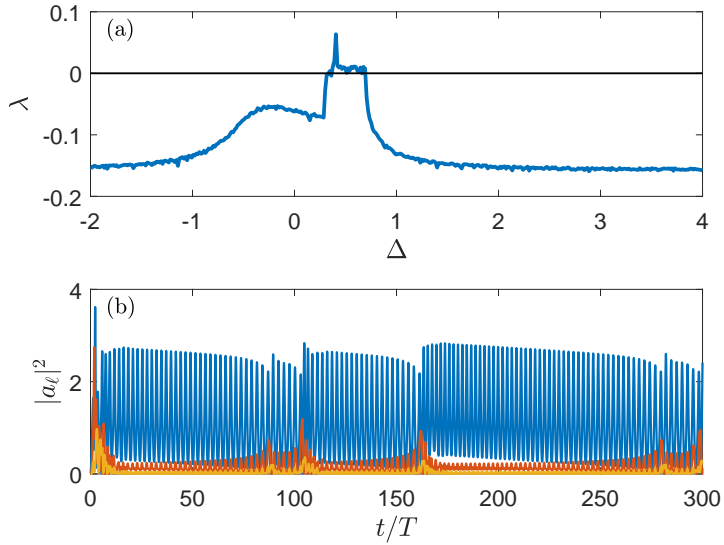}
\caption{(\textbf{a}) Lyapunov exponent of the `stationary' solution of the driven dissipative BH trimer as a function of the detuning $\Delta$.  (\textbf{b}) System dynamics for $\Delta=0.406$. Shown are the squared amplitudes of the local oscillators as the function of time.}
\label{Fig3}
\end{figure}

We repeated calculations for the BH trimer. The dynamics
of the trimer is controlled by the following set of equations of motion
\begin{align}
    & i\dot{a}_1=(-\Delta+U|a_1|^2)a_1-\frac{J}{2}a_2+ \frac{\Omega}{2} \nonumber \\
    & i\dot{a}_2= (-\Delta+U|a_2|^2)a_2-\frac{J}{2}(a_1+a_3)\nonumber \\
    & i\dot{a}_3=(-\Delta+U|a_3|^2)a_2-\frac{J}{2}a_2-i\frac{\gamma}{2}a_3
\end{align}
In the trimer, the new feature is that the above-discussed limit cycle bifurcates into a chaotic attractor, see the upper panel in Fig.~\ref{Fig3},  which shows the Lyapunov exponent of the steady-state solution as the function of the control parameter $\Delta$. An example of this `stationary' solution is given in the lower panel in  Fig.~\ref{Fig3} for $\Delta=0.406$ where the Lyapunov exponent is maximal.

\section{Summary}
\label{sec5}

We showed that attractor bifurcations in the driven dissipative BH system can be well observed in a laboratory experiment with super-conducting circuits already for the value of the effective Planck constant $\hbar=1/4$. In the laboratory experiment, the value of this effective constant is determined by the ratio of the interaction constant $U$ (nonlinearity of the transmon spectrum) to the Rabi frequency $\Omega$, which is proportional to the amplitude of the microwave field. Clearly, the larger the Rabi frequency is, the more photons are simultaneously present in the system. The presented in this work results indicate that the quantum BH dimer reproduces the dynamics of the classical BH dimer when the mean number of photons is of the order of 10.

To conclude, we would like to briefly comment on the experiments where the detuning $\Delta$ is monotonically swept in time. In the present work, we restricted ourselves by the case where $\Delta$ is swept in the negative direction.  If the sweeping direction is inverted,  the result may strongly deviate from that shown in Fig.~\ref{Fig1}(a) due to quantum hysteresis. Within the classical approach, the positive sweeping populates the other attractor whose basin {\em excludes} the point ${\bf a}=0$ \cite{Muraev2022}. In the quantum approach, however, this attractor is a metastable state with a finite lifetime. Thus, unlike the case of negative sweeping, the result of a quasi-adiabatic passage in the positive direction strongly depends on the sweeping rate $d\Delta/dt$.

\section{Acknowledgement}
This work has been supported by Russian Science Foundation through grant N19-12-00167.

\bibliography{mybib}

\end{document}